\documentstyle[12pt]{article}
\newcommand\be{\begin{equation}}
\newcommand\ee{\end{equation}}
\newcommand\ba{\begin{eqnarray}}
\newcommand\ea{\end{eqnarray}}

\newcommand{\cl}{\centerline}
\def\undertilde#1{{\baselineskip=0pt\vtop{\hbox{$#1$}
\hbox{$\scriptscriptstyle\sim$}}}{}}
\begin{document}
\begin{titlepage}
\setlength{\textwidth}{5.0in}
\setlength{\textheight}{7.5in}
\setlength{\parskip}{0.0in}
\setlength{\baselineskip}{18.2pt}
\hfill
{\tt HD-THEP-02-19}
\begin{center}
{\large{\bf Lagrangean Approach to Hamiltonian Gauge Symmetries and the Dirac Conjecture}}\par
\vskip 0.3cm
\begin{center}
{Heinz J. Rothe}\par
\end{center}
\vskip 0.3cm
{Institut f\"ur Theoretische Physik}\par
{Universit\"at Heidelberg, Philosophenweg 16, D-69120 Heidelberg, Germany}\par
\end{center}
\cl{\today}
\vskip 0.2cm
\begin{center}
{\bf ABSTRACT}
\end{center}
\begin{quotation}
\noindent
Using well known Lagrangean techniques for uncovering the gauge symmetries of a Lagrangean, we derive the transformation laws for the phase space variables corresponding to local symmetries of the Hamilton equations of motion. These 
transformation laws are shown to coincide with those derived by  
Hamiltonian methods based on the Dirac conjecture. The connection between the Lagrangean and Hamiltonian approach is illustrated for  
first class systems involving one primary constraint.
\noindent
\end{quotation}
\end{titlepage}

\newpage
It has been conjectured a long time ago by Dirac [1] that local symmetries of the Hamilton equations of motion are intimately tied to the existence of first class constraints, and that the generators for the infinitessimal symmetry transformations are given by the full set of first class constraints. The 
problem of confirming this conjecture has been addressed by several authors [2-8]. In this paper we shall present an alternative approach to this 
problem, based on purely Lagrangean techniques.
 
On the Lagrangean level there exists a well known method for uncovering the local symmetries of a Lagrangean [9,10]. This method leads to gauge transformations which are parametrized by a set of arbitrary functions of time, whose number equals the number of so called "gauge identities" generated by the Lagrangean algorithm. In this letter we shall make use of purely Lagrangean methods 
to derive the transformation laws for the coordinates, momenta, and Lagrange multipliers, which correspond to local symmetries of the Hamilton 
equations of motion. In particular we will show that the transformation laws extracted from the gauge identities are precisely those derived within the Hamiltonian formalism based on the Dirac conjecture. The merit of our approach is that it 
is simple, and that its connection with Hamiltonian methods based on the Dirac conjecture is very transparent. The basic 
ingredients going into the proof will be illustrated for the case of  
first class systems exhibiting one primary constraint.

Consider a system whose dynamics is described by a "singular" Lagrangean 
$L(q,\dot q)$ leading to a set of primary constraints $\{\Omega_\ell\}$. 
Such constraints, following 
alone from the definition of the canonical momenta, must necessarily be present 
if the theory is to exhibit a local symmetry. As 
shown by Dirac [1] the corresponding Hamilton equations of motion are given by 
($i = 1,2,\cdot\cdot\cdot,n$)
\be
\dot q_i = \frac{\partial H}{\partial p_i} + 
\sum_\ell v^\ell\frac{\partial\Omega_\ell}{\partial p_i}
\ee
\be
\dot p_i = -\frac{\partial H}{\partial q_i} - \sum_\ell v^\ell\frac{\partial\Omega_\ell}{\partial q_i}\ ,
\ee
together with the primary constraints
\be
\Omega_\ell(q,p) = 0 \ ,
\ee
where $H(q,p)$ is the canonical Hamiltonian evaluated on the primary surface defined by (3), and where the $v^\ell$'s are undetermined velocities playing the role of Lagrange multipliers.  
These equations follow from a variation principle $\delta S_T = 0$, where 
\be
S_T = \int dt\ \left[\sum^n_{i=1}p_i\dot q_i-H(q,p)-\sum_\ell v^\ell\Omega_\ell\right] \ee
is the "total" action, and the variations are taken with respect to $q_i$, 
$p_i$ and $v$.
 
We now notice that equations (1) and (2) are identical with the {\it Euler-Lagrange} equations of motion derived 
from the first order Lagrangean\footnote{In the following we shall apply 
standard Lagrangean techniques to this first order Lagrangean. They are not to be confused with the symplectic approach discussed in [11].}
\be
L(Q,\dot Q) = \sum^{2n}_{\alpha=1} a_\alpha(Q)\dot Q_\alpha - H_T(Q)\ ,
\ee
with $H_T$,  the "total Hamiltonian, defined by 
\be
H_T = H(Q)+\sum_\ell Q_{2n+\ell}\Omega_\ell(Q) \ ,
\ee
where
$Q_i = q_i$, $Q_{n+i} = p_i$, $Q_{2n+\ell} = v^\ell$  ($i = 1,2,\cdot\cdot\cdot, n$). Note that $H$ and $\Omega_\ell$ only depend on $Q_1,\cdot\cdot\cdot,Q_{2n}$. The non-vanishing $a_\alpha$'s are 
given by $a_i = Q_{n+i}$. The Euler-Lagrange equations of motion read
\be
E^{(0)}_\alpha = 0\ ,
\ee
where $E^{(0)}_\alpha$ is the Euler derivative
\be
E^{(0)}_\alpha = \frac{\partial L}{\partial Q_\alpha} - 
\frac{d}{dt}\left(\frac{\partial L}{\partial\dot Q_\alpha}\right)\ .
\ee
They follow from the variational principle
\be
\delta\int dt\ L(Q,\dot Q) = \sum_\alpha\int dt\ E^{(0)}_\alpha\delta Q_\alpha = 0 \ ,
\ee
where we have dropped a total derivative term in the last integral, and the $\delta Q_\alpha$'s are 
arbitrary variations. Local symmetries of the theory 
correspond to {\it particular} variations $\delta Q_\alpha$ for which (9) 
vanishes {\it without} making use of the equations of motion. 

The merit 
of having recast the variational problem $\delta S_T = 0$ in the above form, is that we can 
make use of well established Lagrangean methods [10] to extract possible local symmetries of the Hamilton equations of motion.  In order to elucidate the basic ideas we shall
consider in the following the case where we are 
dealing with a purely first class theory with only one primary constraint 
$\Omega_1$. 

Consider the Euler derivative (8). From (5) one finds that it has the form 
\be
E^{(0)}_\alpha = \sum_{\beta}F^{(0)}_{\alpha\beta}(Q)\dot Q_\beta 
+K^{(0)}_\alpha(Q)\ ,
\ee
where ($\alpha,\ \beta = 1,\cdot\cdot\cdot,2n+1$)
\be
F^{(0)}_{\alpha\beta}(Q) = \partial_\alpha a_\beta - \partial_\beta a_\alpha \ ,
\ee
\be
K^{(0)}_\alpha(Q) = -\frac{\partial H_T}{\partial Q_\alpha}\ , 
\ee
and
\be
H_T = H(Q) + Q_{2n+1}\Omega_1(Q)\ .
\ee
The only non-vanishing components of the $(2n+1)\times (2n+1)$ matrix 
$F^{(0)}_{\alpha\beta}$ are given by 
$F^{(0)}_{n+i,i} = - F^{(0)}_{i,n+i} = 1$ ($i=1,\cdot\cdot\cdot,n$).
Thus the matrix $F^{(0)}_{\alpha\beta}$ has the form
\be
 {\undertilde F^{(0)}} = \left(
\begin{array}{rrr}
0&-{\undertilde{\bf 1}}&{{\bf\vec 0}}\\
{\undertilde{\bf 1}}&\ 0&{\bf{\vec 0}}\\
{\vec 0}&{\vec 0}&0\\
\end{array}\right)
\ee
where ${\undertilde{\bf 1}}$ is an $n\times n$ unit matrix, ${\bf{\vec 0}}$ is the n-component zero column vector, and $\vec 0$ is corresponding row vector. 
We now look for left-zero modes of $F^{(0)}_{\alpha\beta}$. The only zero 
mode is given by
\be
\vec v^{(0)} = (\vec 0,\vec 0,1) \ ,
\ee
where $\vec 0$ is the $n$-component null vector. Hence
\be
\vec v^{(0)}\cdot\vec E^{(0)} = \vec v^{(0)}\cdot\vec K^{(0)} = 
-\Omega_1 \ .
\ee
On the zeroth level of the Lagrangean algorithm we therefore just recover the 
primary constraint. Per construction, it vanishes on the space of solutions 
to the Euler Lagrange equations of motion (7) (i.e., on shell). This constraint 
is only a function of the "coordinates" $Q_1,\cdot\cdot\cdot, Q_{2n}$. Hence 
its time derivative is given by
\be
\dot\Omega_1 = \sum^{2n}_{\alpha=1}\frac{\partial\Omega_1}{\partial Q_\alpha}\dot Q_\alpha\ .
\ee
We next look for further constraints hidden in the equations of motion. To 
this effect we adjoin (17) to the equations of motion and consider the 
vector 
\be
\vec E^{(1)} = \left(
\begin{array}{r}
\vec E^{(0)}\\
-\frac{d\Omega_1}{dt}
\end{array}\right)
= \left(
\begin{array}{c}
\vec E^{(0)}\\
\frac{d}{dt}(\vec v^{(0)}\cdot\vec E^{(0)})
\end{array}\right)
 \ee
which by construction vanishes again on-shell. 
The $2n+2$ components of 
$\vec E^{(1)}$ can again be written in the form 
\be
E^{(1)}_{\alpha_1} = \sum_{\beta}F^{(1)}_{\alpha_1\beta}(Q)\dot Q_\beta 
+K^{(1)}_{\alpha_1}(Q) \ ; \ \alpha_1 = 1,2,\cdot\cdot\cdot, 2n+2 \ ,
\ee
where ${\undertilde F}^{(1)}$ is the rectangular $(2n+2)\times (2n+1)$ matrix
\be
{\undertilde F}^{(1)} = \left(
\begin{array}{c}
{\undertilde F}^{(0)}\\
\vec\chi_1\\
\end{array}\right)
\ee
with $\vec\chi_1$ the row-vector $(-\vec\nabla\Omega_1,0)$, and $\vec K^{(1)} = (\vec K^{(0)},0) 
= (-\vec\nabla H_T,0)$, where
\be 
{\vec\nabla} = (\partial_1,\partial_2,\cdot\cdot\cdot,\partial_{2n}) \ .
\ee
The zero modes of ${\undertilde F}^{(1)}$ include the previous one, augmented by 
a zero (which will merely reproduce the zeroth-level constraint), 
and a new zero mode given by
\be
\vec v^{(1)} = (-\partial_{n+1}\Omega_1,\cdot\cdot\cdot,-\partial_{2n}\Omega_1,
\partial_{1}\Omega_1,\cdot\cdot\cdot,\partial_{n}\Omega_1,
0,1)\ .\ 
\ee
One then readily verifies that
\be
\vec v^{(1)}\cdot \vec E^{(1)} = \vec v^{(1)}\cdot \vec K^{(1)} 
= \{H_T,\Omega_1\} \ ,
\ee  
where the Poisson bracket is taken with respect to $q$ and $p$ (In the following Poisson brackets are always understood to be taken with respect to $q$ and $p$).  Since 
in the present case we only have one primary constraint, this Poisson bracket 
ist just $\{H,\Omega_1\}$. It is only a function of the $q_i$'s and $p_i$'s, 
i.e., of $Q_1,\cdot\cdot\cdot, Q_{2n}$. 
Let us assume, e.g., that it does not vanish on the primary surface.
\footnote{In order to demonstrate the general method we have to consider 
the case where at least one further constraint is generated.}
Then
\be
-\Omega_2 \equiv \{H,\Omega_1\} = \vec v^{(1)}\cdot\vec E^{(1)}
\ee
is a secondary constraint, vanishing (by construction) on shell.  
We now proceed with the Lagrangean algorithm by adjoining the time 
derivative of $\Omega_2$ to the equations of motion. We are therefore led 
to consider the vector 
\be
\vec E^{(2)} = \left(
\begin{array}{c}
\vec E^{(0)}\\
-\dot\Omega_1\\
-\dot\Omega_2
\end{array}\right)
= \left(
\begin{array}{c}
\vec E^{(0)}\\
\frac{d}{dt}(\vec v^{(0)}\cdot\vec E^{(0)})\\
\frac{d}{dt}(\vec v^{(1)}\cdot\vec E^{(1)})
\end{array}\right) \ .
\ee 
$\vec E^{(2)}_{\alpha_2}$ can again be written in the form
\be
E^{(2)}_{\alpha_2} = \sum_{\beta}F^{(2)}_{\alpha_2\beta}(Q)\dot Q_\beta 
+ K^{(2)}_{\alpha_2}
\ee
where $\alpha_2 = 1,2,\cdot\cdot\cdot,2n+3$, and 
\be
{\undertilde F}^{(2)} = \left(
\begin{array}{c}
\ \ {\undertilde F}^{(0)}\\
\vec\chi_1\\
\vec\chi_2\\
\end{array}\right)
\ee
with $\vec\chi_2$ the row-vector $(-\vec\nabla\Omega_2,0)$. The non-vanishing components 
of the $2n+3$ component vector $\vec K^{(2)}$ are again those of $\vec K^{(0)}$.  ${\undertilde F}^{(2)}$ possesses, besides the previous eigenvectors, augmented by 
an apropriate number of zeros, a new eigenvector having again the generic structure
\be
\vec v^{(2)} = 
(-\partial_{n+1}\Omega_2,\cdot\cdot\cdot,-\partial_{2n}\Omega_2,
\partial_{1}\Omega_2,\cdot\cdot\cdot,\partial_{n}\Omega_2,0,0,1)
\ee
Because of this generic structure, it follows that
\be
\vec v^{(2)}\cdot \vec E^{(2)} =\vec v^{(2)}\cdot \vec K^{(2)} = 
\{H_T,\Omega_2\}\ ,
\ee
or
\be
\vec v^{(2)}\cdot \vec E^{(2)} = \{H,\Omega_2\}+Q_{2n+1}\{\Omega_1,\Omega_2\}\ .
\ee
Note that now the RHS depends on the Lagrange multiplier $v = Q_{2n+1}$, if the Poisson bracket of the constraints (always taken with respect to $q$ and $p$) does not vanish strongly. If we were to adjoin the time derivative of $\vec v^{(2)}\cdot \vec E^{(2)}$ to the equations of motion, we would
be led to an $\vec E^{(3)}$, and corresponding rectangular matrix 
${\undertilde F}^{(3)}$  which possesses no further zero modes. A further constraint can however be generated if $\{\Omega_1,\Omega_2\}$ vanishes on the subspace defined by 
$\Omega_1=\Omega_2 = 0$, i.e., if 
\be
\{\Omega_1,\Omega_2\} = \sum^2_{\ell=1}C^\ell_{12}\Omega_\ell\ .
\ee
In this case we can construct a further function of only $Q_1,\cdot\cdot\cdot,Q_{2n}$ (i.e., which does not depend on the Lagrange multiplier $v$), which vanishes on shell. This function is given by
\footnote{Note that the constraints are generated precisely in the same systematic way as given by the Dirac 
algorithm. There exists therefore a complete correspondence between the Lagrangean and Dirac algorithm.}
\ba 
-\Omega_3 \equiv \{H,\Omega_2\} &=& \vec v^{(2)}\cdot\vec E
^{(2)} -Q_{2n+1}\{\Omega_1,\Omega_2\}\cr
 &=& \vec v^{(2)}\cdot\vec E^{(2)} + Q_{2n+1}\sum^2_{\ell=1}C^\ell_{12}
(\vec v^{(\ell-1)}\cdot\vec E^{(\ell-1)})\ ,
\ea
where we have made use of (30), (16) and (24). If $\Omega_3$ is not a linear combination 
of $\Omega_1$ and $\Omega_2$ then we construct $\vec E^{(3)}$ by adjoining the 
time derivative of $\Omega_3$ to the equations of motion. Proceeding in this way we (possibly) generate successively new constraints given by
\be
-\Omega_{\ell+1} = \vec v^{(\ell)}\cdot \vec E^{(\ell)} - 
Q_{2n+1}\sum^\ell_{k=1}C^k_{1\ell}\Omega_k \ .
\ee
where the $C^k_{1\ell}$ are defined by 
\be
\{\Omega_1,\Omega_{\ell}\} = \sum^\ell_{k=1}C^{k}_{1\ell}(q,p)
\Omega_{k} \ .
\ee
Note that the iterative process allows us to express the RHS of (33) 
in terms of 
scalar products $\vec v^{(n)}\cdot\vec E^{(n)}$ with $n \le \ell$. The algorithm will come to a halt once no new constraint is generated. If $\vec v^{(M)}$ is the last zero mode generated by the iterative procedure, then, for a 
first class system, this will happen when $\vec v^{(M)}\cdot \vec E^{(M)}$ is a linear combination of the constraints $\Omega_1,\cdot\cdot\cdot,\Omega_M$. We are thus left with a so-called "gauge identity",
\be
\vec v^{(M)}\cdot\vec E^{(M)} - \sum^M_{\ell=1}K^\ell_M\Omega_\ell \equiv 0 \ .
\ee
The $Q$-dependent coefficients $K^\ell_M$ can be related to the Poisson brackets $\{\Omega_1,\Omega_M\}$ and $\{H,\Omega_M\}$ as follows. Because of the 
generic structure of the eigenvectors 
\be
\vec v^{(\ell)} = (-\partial_{n+1}\Omega_\ell,\cdot\cdot\cdot,-\partial_{2n}\Omega_\ell,
\partial_{1}\Omega_\ell,\cdot\cdot\cdot,\partial_{n}\Omega_\ell,0,
\cdot\cdot\cdot,0,1)\ \ ; \ \ell > 1
\ee
we have that 
\be
\vec v^{(M)}\cdot\vec E^{(M)} = \{H_T,\Omega_M\} \ ,
\ee
or
\be
\vec v^{(M)}\cdot\vec E^{(M)} = \{H,\Omega_M\} + v\{\Omega_1,\Omega_M\} \ . 
\ee
Furthermore, the constraints generated by the Lagrangean algorithm are identical with those generated successively by the Dirac algorithm. If these 
constraints are first class, then, in particular,
\be
\{\Omega_1,\Omega_M\} = \sum^M_{\ell=1}C^\ell_{1M}\Omega_\ell \ ,
\ee
and (35) and (38) imply that
\be
\{H,\Omega_M\} =  \sum^M_{\ell=1}h^\ell_{M}\Omega_\ell \ .
\ee
Hence
\be
K^\ell_M = h^\ell_M + vC^\ell_{1M} \ .
\ee
Note that in general $h^\ell_M$ and $C^\ell_{1M}$ depend on $Q_1,\cdot\cdot\cdot,Q_{2n}$. 
The constraints appearing in (35) can all be expressed in terms of 
$\vec v^{(\ell)}\cdot\vec E^{(\ell)}$ with $\ell = 1,\cdot\cdot\cdot,M-1$. It 
is therefore evident that the gauge identity (35) can be written in the form
\be
\vec v^{(M)}\cdot\vec E^{(M)} + \sum^{M-1}_{\ell=0}
\rho^{(M)}_{\ell}(Q) (\vec v^{(\ell)}\cdot\vec E^{(\ell)})\equiv 0 \ .
\ee
Because of the iterative way in which the $\vec E^{(\ell)}$ are computed, 
and the generic structure of the eigenvectors (36), this expression can be 
reduced step by step to the form
\be
\sum^{2n+1}_{\alpha=1}\sum^M_{\ell=0}\eta^{(M)}_{\ell}(v^{(\ell)}_\alpha E^{(0)}_\alpha) 
\equiv 0 \ ,
\ee
where
\be
\eta^{(M)}_{\ell} = \eta^{(M)}_{\ell}\left(Q;\frac{d}{dt},
\cdot\cdot\cdot,\frac{d^{M-\ell}}{dt^{M-\ell}}\right) \ .
\ee
The $\eta^{(M)}_{\ell}$ involve linear combinations of time derivatives up to $(M-\ell)$'th order. Multiplying the 
gauge identity (43) by an arbitrary function of time $\alpha (t)$,  
integrating the expression over time, and making 
a sufficient number of partial integrations (dropping total derivative terms, assuming that $\alpha(t)$ and time derivatives thereof up to $M$'th order vanish at the boundaries of integration), one is led to an identity of the form  
\be
\int dt \sum^M_{\ell=0}\epsilon^\ell\left(\sum^{2n+1}_{\alpha=1}
v^{(\ell)}_\alpha E^{(0)}_\alpha\right) \equiv 0 \ .
\ee
The $\epsilon^\ell$'s depend on $Q$ and time derivatives of $Q$, as well as 
on the arbitrary parameter $\alpha(t)$ and time derivatives thereof.
For infinitessimal $\alpha(t)$ expression (45) has the form
\be
\sum_\alpha\int dt\ E^{(0)}_\alpha \delta Q_\alpha \equiv 0 \ ,
\ee
where
\be
\delta Q_\alpha = -\sum^M_{\ell=0}\epsilon^\ell v^{(\ell)}_\alpha \ .
\ee
The minus sign has been included for convenience. 
The integral (46) is just the variation of the 
action induced by the transformation $Q_\alpha\to Q_\alpha+\delta Q_\alpha$ 
(see eq.(9)).
But because of the generic structure of the eigenvectors (36), we have that
(recall that $Q_i = q_i, Q_{n+i} = p_i, Q_{2n+1} = v; \
 i = 1,2,\cdot\cdot\cdot,n$)
\ba
\delta q_i &=& 
\sum^M_{\ell=1}\epsilon^\ell\frac{\partial\Omega_\ell}{\partial p_i} = 
\sum^M_{\ell=1}\epsilon^\ell \{q_i,\Omega_\ell\}\ ,\cr
\delta p_i &=& 
-\sum^M_{\ell=1}\epsilon^\ell\frac{\partial\Omega_\ell}{\partial q_i} = 
\sum^M_{\ell=1}\epsilon^\ell \{p_i,\Omega_\ell\}\ ,\cr
\delta v &=& -\epsilon^0 \ .
\ea
These expressions have precisely the form following from the Dirac conjecture. In fact, the integral (45) can also be cast into a Hamiltonian version. To this effect we notice that it follows from (10), (12) and (14), and the structure 
of the eigenvectors (36) that
\be
\sum^{2n+1}_{\alpha=1}v^{(\ell)}_\alpha E^{(0)}_\alpha = 
\frac{d\Omega_{\ell}}{dt} + \{H_T,\Omega_{\ell}\}\ ; \ \ell \ge 1 \ .
\ee
Hence (45) can also be written in the form 
\be
\int dt\ \left[\sum^M_{\ell=1}\epsilon^\ell\left(\dot\Omega_\ell + \{H_T,\Omega_\ell\}\right)+\delta v\Omega_1\right] \equiv 0 \ ,
\ee
where we have also made use of the identification $\epsilon^0 = -\delta v$.  
This is precisely the form of the variation of the action (4), with the variations $\delta q_i$ and $\delta p_i$ given by (48). Indeed, 
for the case of only one primary constraint, the variation (4), written in these variables, has the form 
\be
\sum_i\int dt \left[\left(\dot q_i-\frac{\partial H}{\partial p_i}
-v\frac{\partial\Omega_1}{\partial p_i}\right)\delta p_i +\left(-\dot p_i- \frac{\partial H}{\partial q_i}
-v\frac{\partial\Omega_1}{\partial q_i}\right)\delta q_i
-\delta v\Omega_1\right] = 0 \ ,
\ee
which, for $\delta q_i$, $\delta p_i$, and $\delta v$ given by (48), then becomes the following {\it identity}:
\be
\int dt\left(\frac{d\epsilon^\ell}{dt}\Omega_\ell -\epsilon^\ell\{H,\Omega_\ell\} 
-v\epsilon^\ell\{\Omega_1,\Omega_\ell\}- 
\delta v\Omega_1\right) \equiv 0 \ .
\ee
This is nothing but Eq. (50), after making a partial integration. 
If we are dealing with a first class system, then 
\be
\{H,\Omega_\ell\} = \sum_k h^k_\ell\Omega_k \ ,
\ee
\be
\{\Omega_1,\Omega_\ell\} = \sum_k C^k_{1\ell}\Omega_k \ .
\ee
Thus (52) becomes
\be
\int dt \left[\left(\frac{d\epsilon^\ell}{dt} -
\epsilon^{\ell'}(h^\ell_{\ell'}+vC^\ell_{1\ell'})\right)\Omega_{\ell}-
\delta v\Omega_1)\right] \equiv 0 \ .
\ee
Hence the $\epsilon^\ell$'s and $\delta v$, determined by the Lagrangean algorithm, should be solutions to
\be 
\frac{d\epsilon^a}{dt} - \epsilon^\ell\left(h_\ell^a
+vC^a_{1\ell}\right) = 0 \ ; \ (a=2,3)\ ,
\ee
\be 
\delta v = \frac{d\epsilon^1}{dt} - \epsilon^\ell\left(h_\ell^1
+vC^1_{1\ell}\right)\ .
\ee
These equations can be solved iteratively [7,12]. The solution involves one arbitrary function of time (corresponding to {\it one} primary 
constraint). 

We now verify explicitely, for the case of an arbitrary 
first class system with one primary and two secondary constraints, that the parameters $\epsilon^\ell$ in (45) determined by the Lagrangean algorithm, are indeed solutions to Eqs. (56). For such a system the  
gauge identity will be generated at the third level of the Lagrangean algorithm. Independent of the specific form of the Hamiltonian one finds, proceeding in the manner described above, that the gauge identity (42) reads
\ba
\vec v^{(3)}\cdot\vec E^{(3)} &+& K^1_3(\vec v^{(0)}\cdot\vec E^{(0)}) 
+K^2_3(\vec v^{(1)}\cdot\vec E^{(1)})\cr
&+&K^3_3\left[\vec v^{(2)}\cdot\vec E^{(2)}+ v\sum^2_{\ell=1}
C^\ell_{12}(\vec v^{(\ell-1)}\cdot\vec E^{(\ell-1)})\right] \equiv 0 \ ,
\ea
where $K^\ell_3$ is given by (41) with $M=3$.
Making use of the generic structure of the eigenvectors (36), and of the recursive 
construction of $\vec E^{(\ell)}$, one then verifies after some algebra that the gauge identity can be recursively reduced to the following form, 
\ba
v^{(3)}_\alpha E^{(0)}_\alpha &+&(K^3_3+\frac{d}{dt})
(v^{(2)}_\alpha E^{(0)}_\alpha)\cr
&+& (K^2_3 + vK^3_3C^2_{12} + K^3_3\frac{d}{dt} + \frac{d}{dt} vC^2_{12} 
+ \frac{d^2}{dt^2})(v^{(1)}_\alpha E^{(0)}_\alpha)\cr
&+& [K^1_3+K^2_3\frac{d}{dt} + K^3_3\frac{d^2}{dt^2}+
vK^3_3 (C^1_{12}+C^2_{12}\frac{d}{dt}) + \frac{d^3}{dt^3}\cr
&+& \frac{d}{dt}v(C^1_{12}+C^2_{12}\frac{d}{dt})](v^{(0)}_\alpha E^{(0)}_\alpha) \equiv 0 \ ,
\ea
where a summation over $\alpha = 1,2,\cdot\cdot\cdot,7$ is understood. The time derivatives act all the way to the right.
By multiplying this expression with an arbitrary function of time $\epsilon(t)$, and integrating the expression over time, dropping a total derivative, we 
find, upon making use of (49), that it can be written in the form (50), where
\ba
\epsilon^3 &=& \epsilon(t) \ ,\cr
\epsilon^2 &=& K^3_3\epsilon-\dot\epsilon\ ,\cr
\epsilon^1 &=& \ddot\epsilon-\frac{d}{dt}(K^3_3\epsilon)-vC^2_{12}\dot\epsilon+K^2_3\epsilon +vK^3_3C^2_{12}\epsilon \ .\cr
\delta v &=& \dot\epsilon^1+vC^1_{12}\dot\epsilon-K^1_3\epsilon-
vK^3_3C^1_{12}\epsilon \ .
\ea
One can now easily verify that (60) are indeed solutions to (56) and (57), i.e., to
\be
\dot\epsilon^a-\sum^3_{\ell=1}\epsilon^\ell K^a_\ell = 0 \ ; \ (a=2,3)\ ,
\ee
\be
\dot\epsilon^1-\sum^3_{\ell=1}\epsilon^\ell K^1_\ell - \delta v = 0 \ .
\ee
Setting $\epsilon^3 = \epsilon(t)$, and making use of the fact that, because 
of the way the constraints have been generated by the Lagrangean algorithm, 
$h^3_1 = C^3_{11} = C^3_{12} = 0$, and $h^2_1 = h^3_2 = -1$ in (56) and (57), one can solve the above equations recursively, starting with equation (61) for $a=3$.  The solutions are given, as expected, by (60).

Concluding, we have shown that well established Lagrangean techniques can be used to uncover all the gauge symmetries of the Hamiltonian equations 
of motion for first class systems with one primary constraint, and that there 
is a direct correspondence between each level of the Lagrangean algorithm 
and the Dirac approach to local symmetries based on the Dirac conjecture. This 
analysis can be extended to systems with several primary constraints. A general discussion of gauge symmetries for arbitrary 
systems based on the approach presented in this paper, is deferred to a 
future publication. 

\end{document}